\documentclass[numberedappendix]{emulateapj} 

\usepackage{natbib}
\usepackage{hyperref}
\usepackage{graphicx}
\hypersetup{colorlinks,citecolor=Blue,linkcolor=Black,urlcolor=Blue}
\usepackage{amsmath}
\usepackage{empheq}
\usepackage[usenames,dvipsnames]{color}

\allowdisplaybreaks 

\begin{document}

\title{Chaotic Excitation and Tidal Damping in the GJ\,876 System}  
\author{Abhijit Puranam \& Konstantin Batygin} 

\affil{Division of Geological and Planetary Sciences, California Institute of Technology, Pasadena, CA 91125} 
\email{kbatygin@gps.caltech.edu}
 
\newcommand{\Ham}{\mathcal{H}}
\newcommand{\G}{\mathcal{G}}
\newcommand{\appropto}{\mathrel{\vcenter{\offinterlineskip
\halign{\hfil$##$\cr\propto\cr\noalign{\kern2pt}\sim\cr\noalign{\kern-2pt}}}}}
\newcommand{\Poincare}{{Poincar$\acute{\rm{e}}$}}

\begin{abstract} 

The M-dwarf GJ$\,$876 is the closest known star to harbor a multi-planetary system. With three outer planets locked in a chaotic Laplace-type resonance and an appreciably eccentric short-period Super-Earth, this system represents a unique exposition of extrasolar planetary dynamics. A key question that concerns the long-term evolution of this system, and the fate of close-in planets in general, is how the significant eccentricity of the inner-most planet is maintained against tidal circularization on timescales comparable to the age of the universe. Here, we employ stochastic secular perturbation theory and $N$-body simulations to show that the orbit of the inner-most planet is shaped by a delicate balance between extrinsic chaotic forcing and tidal dissipation. As such, the planet's orbital eccentricity represents an indirect measure of its tidal quality factor. Based on the system's present-day architecture, we estimate that the extrasolar Super-Earth GJ$\,$876$\,$d has a tidal $Q\sim10^{4}-10^{5}$, a value characteristic of solar system gas giants. 

\end{abstract} 

\maketitle

\section{Introduction} \label{sect1}
The study of main-sequence exoplanetary systems began with the detection of the first Hot Jupiter, 51\,Peg\,b, in 1995 (\citealt {mayorqueloz1995}). At the time, the detection of these massive, close-in planets defied the conventional theory of planet formation (\citealt{pollack1996}). The development of planet migration theory (\citealt{goldtremaine1980}), however, helped explain the existence of Hot Jupiters and furthered the understanding of the physical processes that operate concurrently with planet formation\footnote{The extent to which hot Jupiters migrate within their natal disks remains an open question, as both in-situ and ex-situ formation models have been suggested (\citealt{batygin2016}).}(\citealt{lin1996}).

Among the earliest Hot Jupiters to be detected was GJ\,876b, a 2.2\,M$_J$ planet orbiting an M-dwarf at 0.2\,AU with a period of P = 60\,days (\citealt{marcy1998}). Subsequently, GJ\,876\,c, a 0.7\,M$_J$ planet orbiting at 0.13\,AU, was discovered in a 2:1 orbital resonance with 876\,b (\citealt{marcy2001}). This discovery made the GJ\,876 planetary system not only an ideal testing ground for theories of planetary formation and migration, but also an unusual example of a close-in giant planet system (\citealt{fischervalenti2005}). 

Further characterization of the system revealed GJ\,876\,d, a Super Earth with a semi-major axis of 0.02 AU, and an exterior Uranus-mass planet GJ\,876\,e (\citealt{rivera2005,rivera2010}). GJ\,876\,e orbits at 0.35 AU with a period of P = 120 days, making the c-b-e triplet a Laplace-type resonance. Against the backdrop of the planetary population discovered by the Kepler mission, it is evident that rather than a typical member of the galactic planetary census, the GJ\,876 system stands out as a remarkable anomaly worthy of closer examination.

The GJ\,876 system is unique because (1) Jupiter-class planets are generally rare around M-dwarf stars (\citealt{valentifischer2005}) and (2) the c-b-e Laplace resonance exhibits rapid dynamical chaos, unlike the well ordered Laplace resonance of the Galilean moons of Jupiter (\citealt{batdeck2015}). Additionally, the short-period super-Earth GJ\,876\,d exhibits an intriguingly moderate eccentricity of $e_d=0.05-0.15$ (\citealt{correia2010,nelson2016,Trifonov2018,Sarah}). As we show below, any finite eccentricity is at odds with the standard model of tidal dissipation. Accordingly, this study focuses on the origins of this peculiar orbital architecture. Specifically, here we show that chaotic excitations arising from the Laplace-like resonance of the c-b-e chain are instrumental towards explaining the eccentricity of 876\,d, and this behavior yields constraints on the planet's tidal quality factor.

The paper is structured as follows. In section 2, we discuss our numerical experiments and characterize the dynamical behavior of the system. In section 3, we employ stochastic perturbation theory to model and support the results of our numerical experiments. We conclude and discuss our results in section 4.

\begin{table}
\centering
\caption{Adopted orbital fit of the GJ\,876 system\label{tbl-1}}
\begin{tabular}{l c c c c c c}
\tableline
\tableline
 & $M$ ($M_{\odot}$) & $a$ (AU) & $e$ & $\mathcal{M}$ (deg) & $\varpi$ (deg) \\
\tableline
$\star$ &0.37\\
d &2.25$\times$ 10$^{-5}$ &0.02 &0.12 &301.87 &219.10\\
c &7.99$\times$ 10$^{-4}$ &0.13 &0.25  &139.32 &115.96\\
b &2.54$\times$ 10$^{-3}$ &0.21 &0.03 &74.62 &110.86\\
e &4.98$\times$ 10$^{-5}$ &0.35 &0.03 &184.11 &129.56\\

\tableline

\end{tabular}
  \label{tab:myfirsttable}
\end{table}
\section{Numerical Experiments} \label{sect2}
The physical parameters of the GJ\,876 system have been refined continuously since its discovery. For the purposes of this work, we have adopted the recent fit of orbital parameters from \citet{nelson2016}. This fit is based on Doppler velocity measurements and results in an orbital configuration that exhibits long-term stability.

Our numerical simulations utilized the $Symba$ symplectic gravitational dynamics software package, based on the work of \citet{wishol1991,duncan1998}. In order to properly model short-range planet-star interactions, the code has been modified to account for general relativistic effects (\citealt{nobiliroxburgh1986}) and to include tidal effects using the constant time-lag formalism outlined by \citet{mignard1980,hut1981}. In particular, dissipative effects were self-consistently taken into account by introducing velocity-dependent forces into the equations of motion, while ensuring that total (spin plus orbital) angular momentum remains conserved \citep{MardlingLin2002}. For all integrations, we adopted a timestep of $0.1\,$days - approximately 1/20th of the inner-most planet's orbital period. 

\subsection{Tidal Evolution}
In the absence of external gravitational perturbations the tidal evolution of eccentricity occurs on a characteristic timescale (\citealt{yoderpeale1981}):
\begin{equation}
\Big(\frac{1}{e}\frac{de}{dt}\Big)^{-1} = \Big(\frac{1}{n}\Big)\Big(\frac{2}{21}\Big)\Big(\frac{m}{M_\star}\Big)\Big(\frac{a}{R}\Big)^5\Big(\frac{Q}{k_2}\Big)=\tau_e,
\label{taue}
\end{equation}
where $n$ is the mean motion, $a$ is the semimajor axis, $R$ is the radius of the orbiting body, $Q$ is the tidal dissipation quality factor, and $k_2$  is the Love number. We note that in addition to eccentricity damping, tidal dissipation also facilitates semi-major axis decay. However, this process unfolds on a much longer characteristic timescale: $\tau_a = \tau_e/2\,e^2\sim30\,\tau_e$ and can be neglected for the moment (our numerical experiments will treat tidal dissipation self-consistently). 

Expression (\ref{taue}) indicates that in the presence of only tidal forcing, a planet's eccentricity will exhibit nearly-exponential decay. Calculating the true circularization timescale of an exoplanet directly using this expression, however, is abortive because first-principles based estimates of the tidal quality factor remain elusive, with lack of constraints on the planetary composition further aggravating the uncertainty. Instead, the oft-cited estimates tidal $Q$ are based on observations of bodies within our own solar system \citep{goldreichsoter1966}, with no faithful point of comparison for the tidal quality factor of 876\,d, a Super-Earth.
	
To motivate further calculations, let us first consider a scenario where the only relevant forcing on eccentricity comes from tidal dissipation. The dynamics would then be described by equation (1), and the currently observed high eccentricity would be due to a tidal dissipation timescale on the order of the lifetime of the system, about 10 Gyr. Using the mass and orbital parameters from table (1), we may draw upon the exoplanet mass-radius relationship obtained by \citet{weissmarcy2013} to estimate a planetary radius of R $\sim$ 3 $R_{\oplus}$. Then, using a Love number of $k_2$ $\sim$ 0.4 (\citealt{dermott1988,banmurray1992}), we obtain a tidal quality factor on the order of $Q\gtrsim3\times10^6$. This is a value considerably larger than that of solar system gas giants, so either 876\,d is more than an order of magnitude less dissipative than these planets, or other factors must be important to the evolution of 876\,d's eccentricity. Let us consider this alternative.

Our first numerical experiment aims to examine the evolution of GJ\,876\,d's eccentricity in consideration of tidal forcing from its host star and the periodic gravitational perturbations of the giant planets in the system. We do this by considering the system in a nonchaotic regime by removing 876\,e from our calculations. This renders the evolution of planets b and c strictly periodic, because the stochastic evolution of 876\,e defines the chaotic nature of the system (\citealt{batdeck2015}). At the same time, planet e's low mass and significant distance from 876\,d result in negligible direct influence on the evolution of 876\,d's orbit. Thus, the results of these simulations elucidate the behavior of 876\,d's eccentricity in the absence of chaotic effects. 

In our numerical experiments, we used shorter circularization times than those implied by conventional theory, so that many dynamical timescales can be simulated with less computation time. We note that this adjustment is valid as long as the timescales associated with angular momentum exchange among the planets remain much shorter than the circularization timescales we have adopted.

The results of a representative simulation are shown in Figure (1). Clearly, this calculation yields no explanation for the high eccentricity of planet d. That is, in presence of tidal forcing, no significant excitations in eccentricity exist. Instead, the behavior of 876\,d's eccentricity is governed by the exponential decay from tides with small periodic perturbations from the outer planets. Thus, strictly periodic gravitational perturbations from the outer giant planets alone cannot explain the high eccentricity that is observed.

\begin{figure}[t]
	\centering
		\includegraphics[scale=.6]{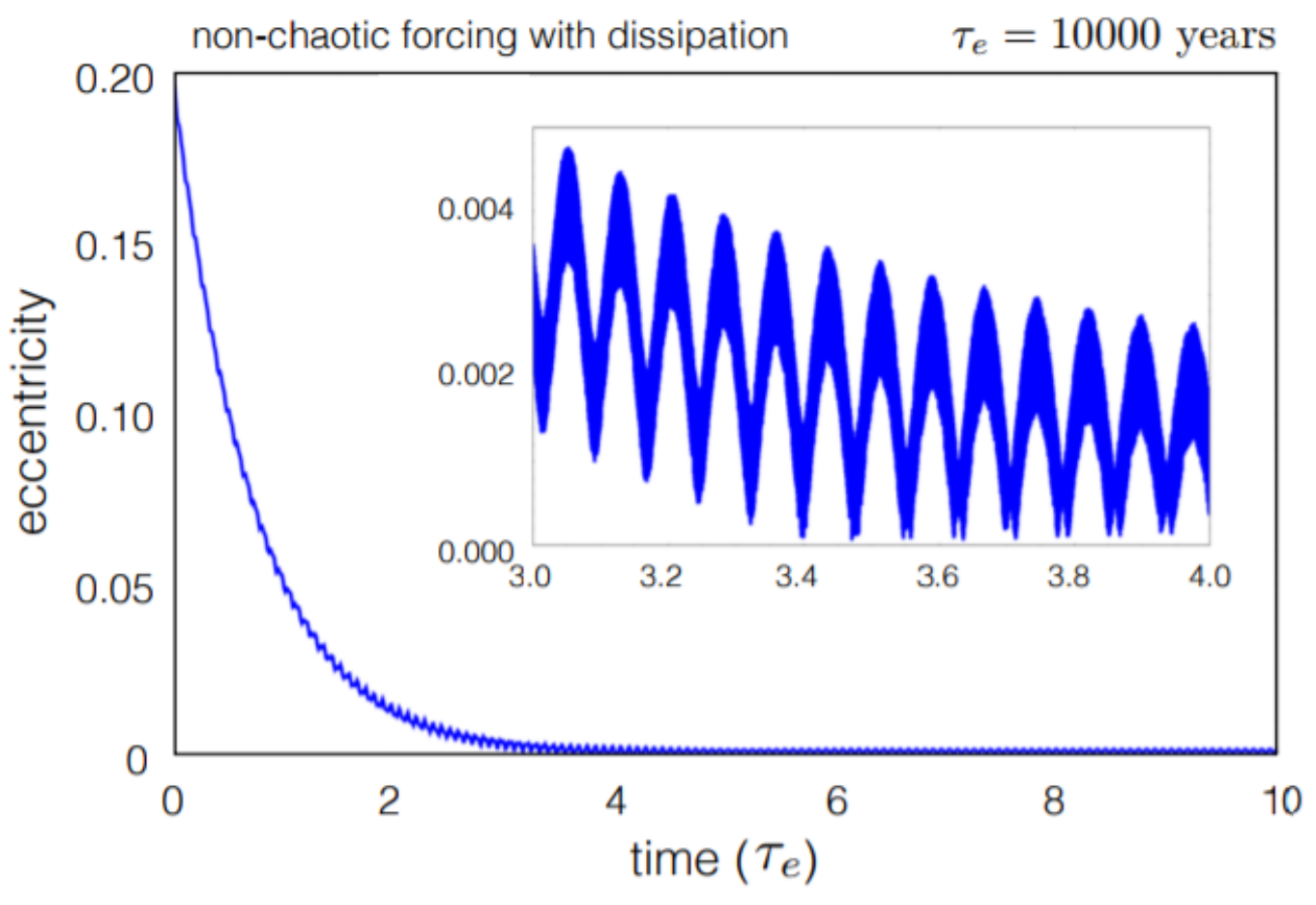}
		\caption{Evolution of GJ\,876\,d's eccentricity subject to tidal dissipation as well as periodic gravitational forcing from planets b and c. The simulation was performed with initial eccentricity of 0.2 and a tidal timescale of $\tau_e=10,000$ years.}
\end{figure}
\subsection{Chaotic Dynamics}
We now consider the full system and its chaotic behavior. Unlike the simulation shown in Figure (1), here we model the full planetary system without considering tidal forces. Our results, shown in Figure (2), demonstrate that chaotic forcing alone is enough to drive eccentricity of 876\,d to the observed value on a timescale of a few million years, much less than the lifetime of the system. 

\begin{figure}[t]
	\centering
		\includegraphics[scale=.6]{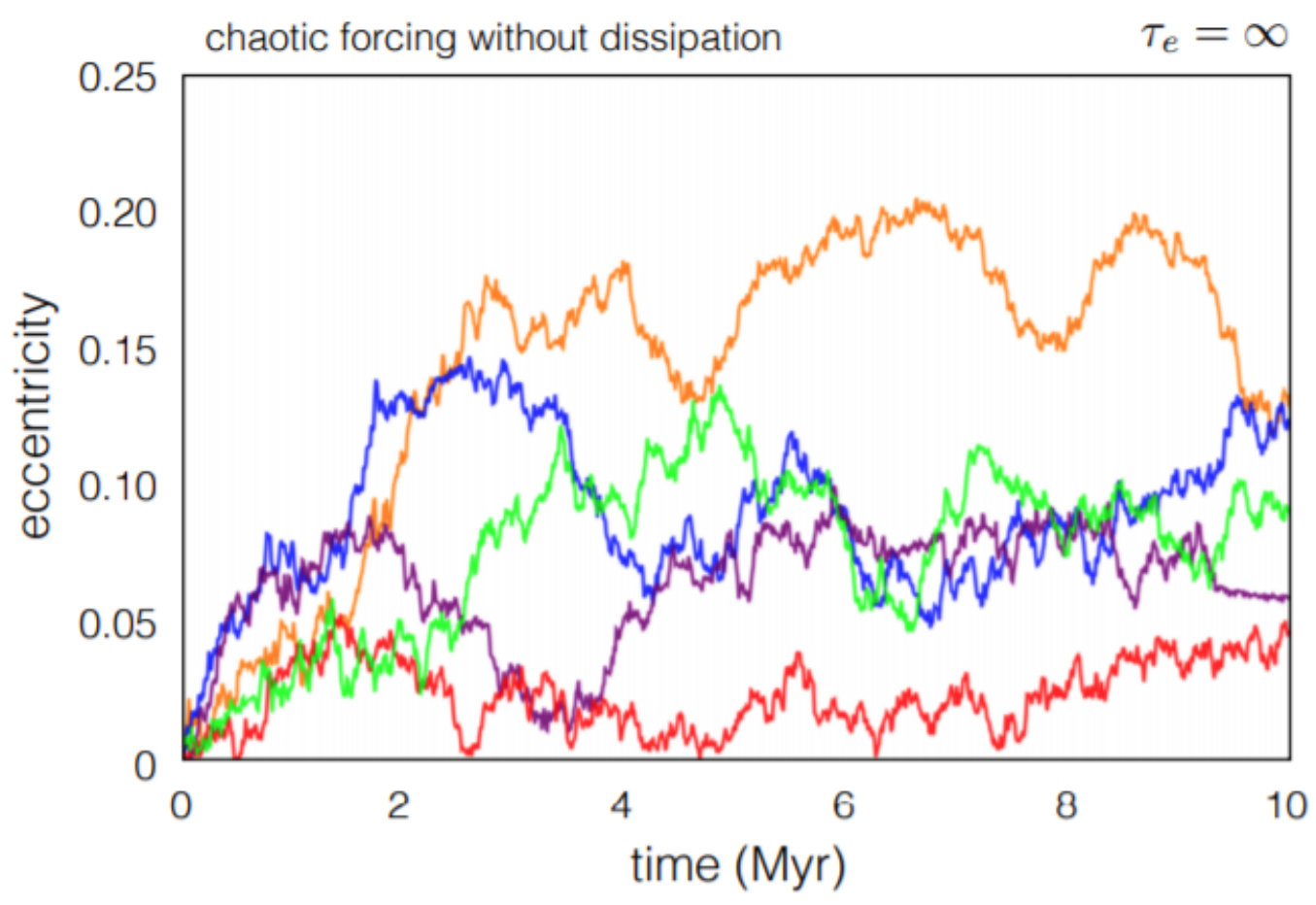}
		\caption{Chaotic evolution of GJ\,876\,d's eccentricity using randomly generated initial mean anomalies. Note that in absence of tidal effects, the chaotic eccentricity excitation timescale is orders of magnitude shorter than the multi-Gyr age of the system.}
	\end{figure}	
	
With this insight into the importance of chaotic forcing, we perform our tidal dissipation numerical experiment again, this time accounting for the presence of 876\,e. The behavior of planet d's eccentricity in this case is shown in Figure (3). While the dominant exponential decay of the eccentricity is similar, the periodic perturbations that manifested in the first experiment are replaced by chaotic excitations in the second experiment. Correspondingly, at sufficiently low eccentricities, these chaotic excitations dominate the behavior over the exponential decay of the tidal forcing.

To quantify this relationship between the tidal and chaotic forcing, we ran a series of simulations sequentially increasing $\tau_e$. A range of twenty different tidal circularization timescales were taken, and for each one the simulation was performed over ten circularization timescales. The chaotic nature of the system is acutely sensitive to initial conditions, making it difficult to extract precise information from the system at different circularization timescales - that is, each simulation is only representative of the true evolution of the system in a statistical sense. To account for these effects, ten simulations were performed at every timescale, with a randomly generated initial mean anomaly for 876\,d's orbit\footnote{Given the purely secular nature of gravitational coupling between planet d and the remainder of the system, changing its mean anomaly has no qualitative impact on the ensuing dynamical evolution.}. The results are shown in Figure (4), where RMS eccentricity of planet d (measured over the last seven circularization timescales) is depicted as a function of $\tau_e$. Clearly, as circularization time is increased, the mean eccentricity attained by planet d also grows. Quantifying this relationship is the focus of the next section.

\section{Analytic Theory} \label{sect3}
\subsection{Stochastic Secular Perturbation Theory}
Numerical experiments presented in the preceding section suggest that the average eccentricity acquired by planet d after a period of chaotic equilibration is a monotonically increasing function of $\tau_e$. To understand this relationship from analytic grounds, in this section we construct a simplified, perturbative model for the inner-most planet's motion. Given the large separation between the semi-major axes of planets d and c (as well as b) and the lack of a low-order commensurability between the orbital periods, it is sensible to assume that the interactions between these objects are secular in nature. Accordingly, in order to obtain a handle on the dynamical evolution of planet d, we employ Lagrange-Laplace perturbation theory (\citealt{md99}).

To second order in the orbital eccentricities, the secular Hamiltonian that governs GJ\,876\,$d$'s evolution reads:
\begin{align}
\mathcal{H} &= \frac{-1}{\sqrt{\mathcal{G} M a }}\frac{\mathcal{G} m'}{4\,a'}\bigg(\frac{a}{a'}\bigg)^{2}\bigg[b_{3/2}^{(1)}\,\Gamma\nonumber \\
&- b_{3/2}^{(2)} e'\, \sqrt{2\,\Gamma} \cos(\xi\,t - \gamma) \bigg]
\label{HLL}
\end{align}
where $b$'s are Laplace coefficients, $\Gamma = 1- \sqrt{1-e^2} \simeq e^2/2$ and $\gamma=-\varpi$ are scaled \Poincare\ action-angle variables corresponding to planet d, while the primed quantities correspond to the perturbing body (i.e. planet c or b). For definitiveness, we assume that $e'$ is nearly constant and the rate of the perturbing planet's pericenter regression, $\xi$, is time-independent. We note that both of these assumptions are justified by numerical simulations and simultaneously apply to planets b and c, since these objects have co-linear apsidal lines (\citealt{correia2010,batdeck2015}).

\begin{figure}[t]
	\centering
		\includegraphics[scale=.6]{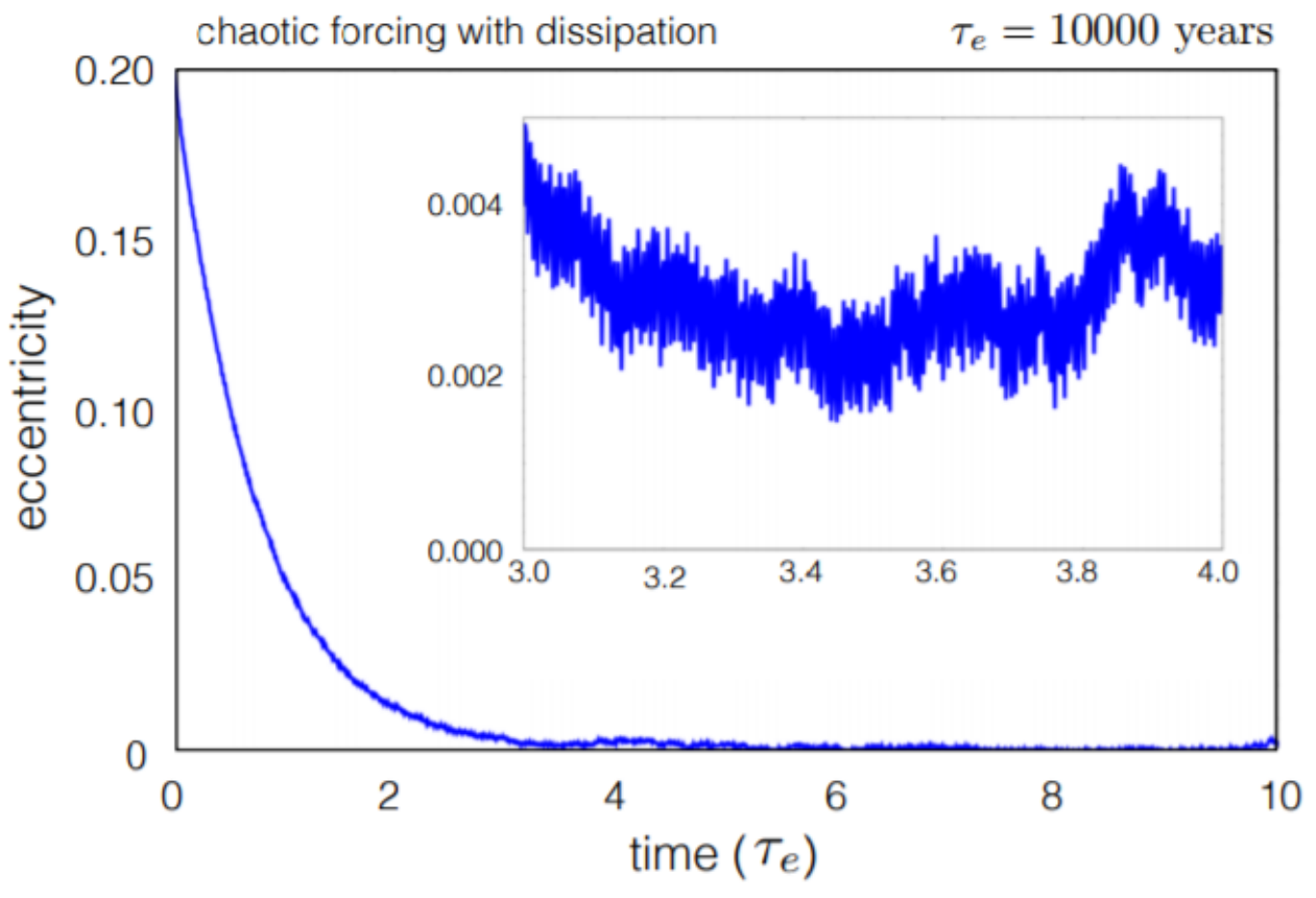}
		\caption{Chaotic Evolution of GJ\,876\,d's eccentricity in presence of tidal dissipation. The simulation was performed at an initial eccentricity of 0.2 with a tidal timescale of 10,000 years.}
	\end{figure}	

To leading order in the semi-major axis ratio, Hamilton's equation for planet d's longitude of perihelion is:
\begin{align}
\frac{d\gamma}{dt} &= \frac{\partial\mathcal{H}}{\partial\Gamma}\simeq -\frac{3}{4} n \bigg(\frac{m'}{M}\bigg) \bigg(\frac{a}{a'}\bigg)^3 + \mathcal{O}\bigg(\frac{a}{a'} \bigg)^4,
\label{precession}
\end{align}
where $n$ is the mean motion of planet d, and we have expanded the Laplace coefficient as a hypergeometric series to leading order. Using equation (\ref{precession}), it is trivial to check that $|d\gamma/dt|\ll \xi$. This disparity in precession frequencies explains why no coherent angular momentum exchange can be seen in Figure (1). That is, resonantly-induced rapid regression of planet b and c's pericenter essentially renders the secular harmonic of Hamiltonian (2) a short-periodic effect that can be averaged over.

Accounting for the tidal decay of the orbital eccentricity (e.g. \citealt{hut1981}), the equation of motion for planet d's angular momentum deficit reads:
\begin{align}
\frac{d\Gamma}{dt} &= -\frac{\partial\mathcal{H}}{\partial\gamma}+\frac{\partial\Gamma}{\partial e}\left(\frac{de}{dt}\right)_{\rm{tide}} \nonumber \\
&\simeq \frac{15}{8\sqrt{2}} n \bigg(\frac{a}{a'}\bigg)^4\, e'\,\sqrt{\Gamma} \sin(\xi\,t) - 2\frac{\Gamma}{\tau_e},
\label{AMD}
\end{align}
where we have neglected the slow evolution of $\gamma$ and adopted $\xi$ as the harmonic circulation frequency. Averaged over timescales much longer than $2\pi/\xi$, chaotic variation in $e'\,\sin(\xi\,t)$, that arises from perturbations due to planet e, can be crudely modeled as a drift-free Weiner process, $\mathcal{W}$ (\citealt{batholman2015}). Under this prescription, equation (\ref{AMD}) reduces to a readily integrable stochastic differential equation:
\begin{align}
d\langle\Gamma\rangle &= 2\sqrt{\mathcal{D} \langle\Gamma\rangle}d\mathcal{W} - 2 \frac{\langle\Gamma\rangle}{\tau_e}dt,
\label{stochastic}
\end{align}
where $\langle\Gamma\rangle$ denotes the time-averaged angular momentum deficit, and $\mathcal{D}$ is an effective diffusion coefficient that encompasses the chaotic nature of the perturbing orbits. 

The standard deviation of the distribution function arising from equation (\ref{stochastic}) is given by
\begin{align}
\sigma_{\langle\Gamma\rangle} &= \sqrt{2\,\langle\Gamma_0\rangle\,\mathcal{D}\,\tau_e \big(\exp(-2t/\tau_e)-\exp(-4t/\tau_e) \big)},
\label{sigmafull}
\end{align}
where $\langle\Gamma_0\rangle\ne0$ is an (arbitrary) initial condition. The function (\ref{sigmafull}) has a global maximum at $t=\log(2)\tau_e/2$, and asymptotically approaches zero as $t\rightarrow\infty$. This long-term behavior is unphysical, and stems from the fact that $\langle\Gamma\rangle=0$ is an absorbing boundary in equation (\ref{stochastic}), while in reality it is reflective (i.e. a path that encounters $\langle\Gamma\rangle=0$ within the framework of equation (\ref{stochastic}) will remain at $\langle\Gamma\rangle=0$ for all subsequent time, whereas real circular orbits can become eccentric\footnote{This unphysical behavior results from decoupling Hamilton's equations from one another, and only manifests as a practical issue at very low eccentricities, where $d\gamma/dt$ can become arbitrarily large.}). Here, we circumvent this artificial limitation simply by evaluating $\sigma_{\langle\Gamma\rangle}$ at its global maximum. Then, the standard deviation of the stochastic process (\ref{stochastic}) becomes
\begin{align}
\big(\sigma_{\langle\Gamma\rangle}\big)_{\rm{max}} &= \sqrt{\langle\Gamma_0\rangle\,\mathcal{D}\,\tau_e /2 }.
\label{sigmafull}
\end{align}

\begin{figure}[t]
	\centering
		\includegraphics[scale=.6]{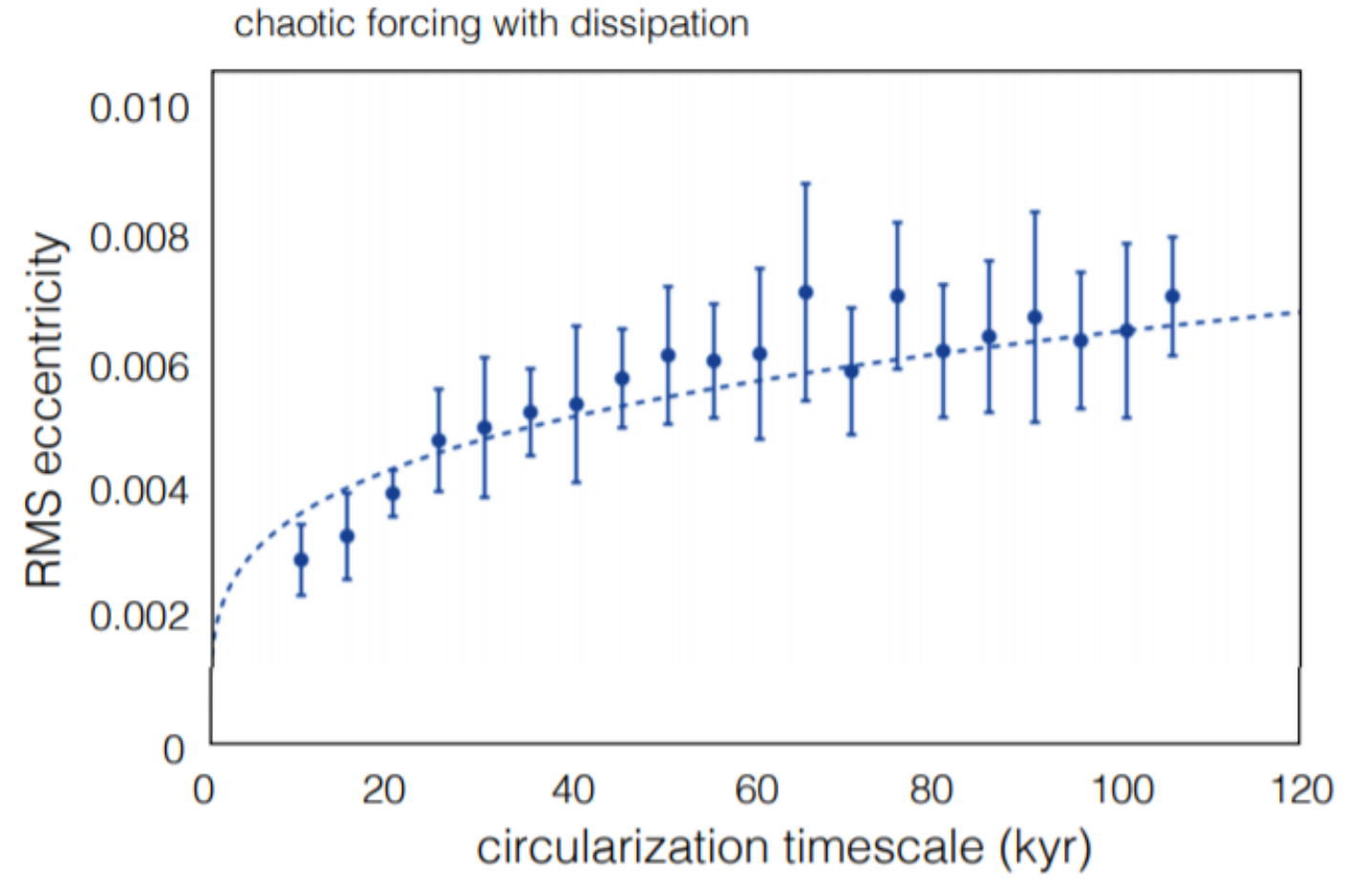}
		\caption{RMS eccentricity of planet d, as a function of tidal decay timescale. The depicted points and error-bars respectively denote the mean and the standard deviation of the RMS eccentricity sampled across ten simulations at each circularization timescale. Our suite of numerical experiments reveals a monotonically increasing relationship between that the root mean square eccentricity of 876\,d and the tidal damping timescale.}
	\end{figure}

The key result of the above analysis is that the average eccentricity of planet d scales as the quarter-root of the circularization timescale:
\begin{align}
\langle e\rangle\appropto\tau_e^{1/4}.
\label{quarterroot}
\end{align}
Note that this scaling was derived exclusively under the assumption of secular perturbations arising from a chaotic companion, and should therefore be applicable even if planet d experiences a limited degree of inward tidal migration over the lifetime of the system. The relationship (\ref{quarterroot}) is shown as a blue dashed line in Figure (4), and clearly provides an adequate match to our suite of numerical simulations, allowing us to extrapolate the existing results to more realistic values of $\tau_e$.
\subsection{A lower bound on tidal $Q$}
The preceding theory can be combined with the results of the numerical experiment described in Section 2.2 to estimate the lower bound of GJ\,876\,d's tidal quality factor. Specifically, within the context of our N-body simulations, the maximum value of the eccentricity attained by planet d over the modeled $10\tau_e$ integration period is well fit by a quarter-root function:
\begin{align}
\langle e\rangle = 9.1\times10^{-4}\bigg(\frac{\tau_e}{\rm{yr}}\bigg)^{1/4}.
\label{maxrelationship}
\end{align}
\begin{figure}[t]
	\centering
		\includegraphics[scale=.59]{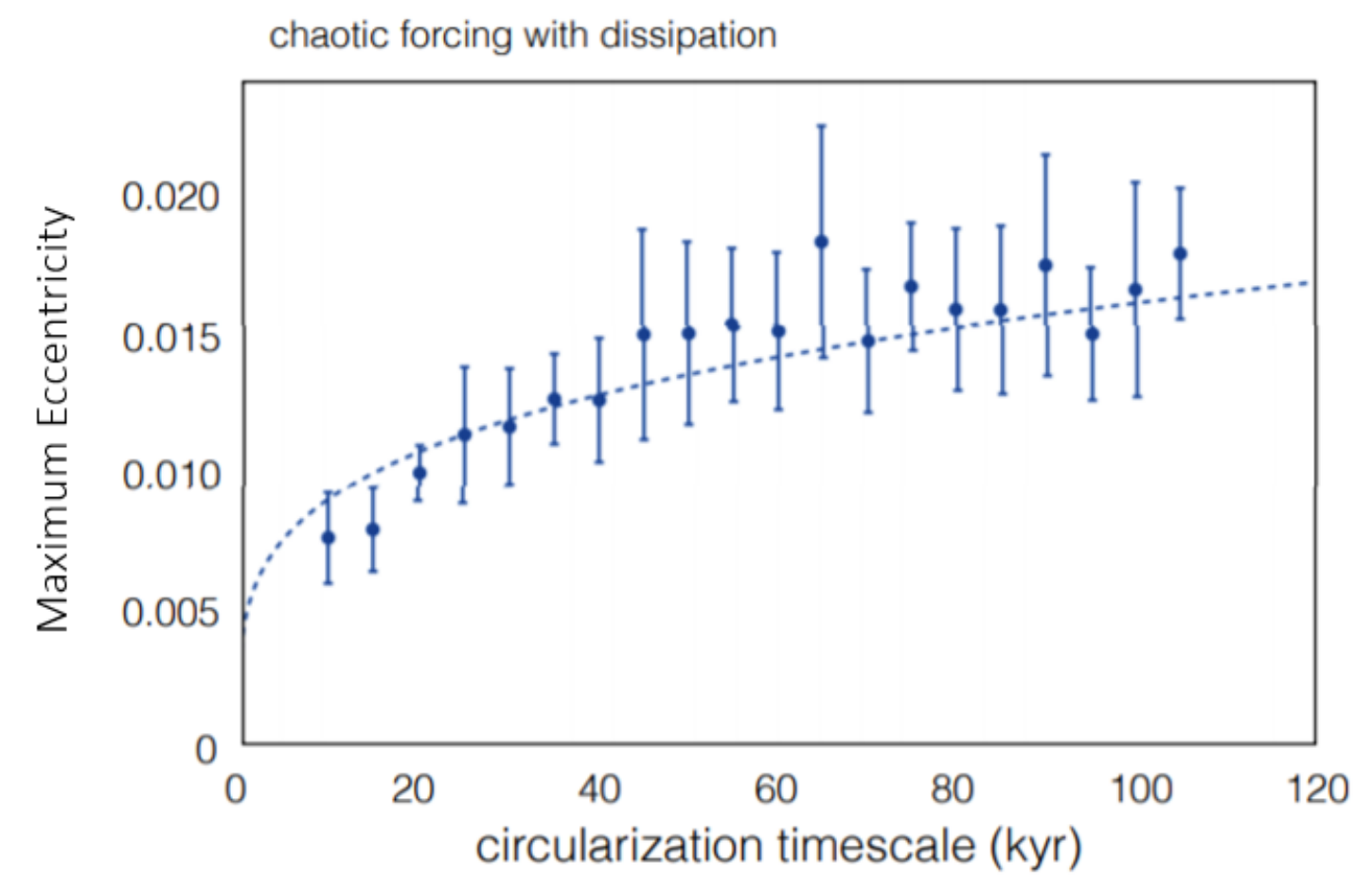}
		\caption{The maximum eccentricity achieved as a function of tidal timescale for our numerical experiment. As in Figure (4), the depicted points and error-bars respectively denote the mean and the standard deviation of the maximal eccentricity sampled across ten simulations at each circularization timescale. The results follow the analytically expected quarter-root relationship indicated by the dashed line.}
	\end{figure}

This relationship, shown in Figure (5), can readily be extrapolated to the observed eccentricity to determine an approximate lower-bound on the tidal circularization timescale of GJ\,876\,d. Using the parameters described in Table (1), we estimate that the maximum circularization timescale is of order $\tau_e\sim3\times10^{8}$ years. Applying this value to equation (1) and assuming a planetary radius of 3\,$R_{\Earth}$ (\citealt{weissmarcy2013}) and a Love number of $k_2=0.4$ as before (\citealt{banmurray1992}; \citealt{dermott1988}), we obtain a lower-bound on the tidal quality factor of order $Q\sim10^5$.
\section{Discussion} \label{sect4}
In this paper, we have examined the origins of the suspiciously finite eccentricity of GJ\,876\,d, and identified the dynamical mechanism that allows for a non-zero eccentricity to be maintained against tidal dissipation. This mechanism relies on the rapid chaotic diffusion of the perturbing planets, and thus requires a short Lyapunov time to operate. This means that the described mechanism is specific to the orbital architecture of the GJ\,876 system, with no dynamical equivalent existing within the Solar System.

The tidal $Q$ we have obtained is in essence a lower bound, and represents the first purely dynamical derivation of the specific dissipation function of a super-Earth. This analysis compliments the recent findings of \citet{morleyetal}, who have derived a $Q\sim10^5$ for GJ436b, a Neptune-mass extrasolar planet. Within an order of magnitude, these values also agree with the values of $Q$ derived for solar system ice-giants (\citealt{md99}).

GJ\,876\,d does not transit its host star, so we were forced to make an assumption about the planetary radius in order to estimate tidal $Q$. In the relevant mass range, the extrasolar radius-mass data used by \citet{weissmarcy2013} indicates that true radius of GJ\,876\,d could feasibly lie in the range of 2 - 3.5 $R_{\Earth}$, although detailed models of \citet{Valencia2007} indicate that the radius could in principle lie below 2 $R_{\Earth}$. In particular, this would lead our lower bound estimate of tidal quality to range from order $10^4$ to order $10^6$. Unfortunately, without direct measurements of the planetary radius and interior density distribution, it is impossible to derive tighter empirical constraints on this value, leaving our fiducial $Q\sim10^5$ estimate of the quality factor highly uncertain.

Additional ambiguity in our calculations arises from the fact that the eccentricity of planet d itself is relatively poorly constrained by the radial velocity data. To this end, the recent analyses of \citet{Trifonov2018,Sarah} favor a somewhat lower eccentricity than that derived by \citet{nelson2016} i.e., $e_d\approx0.06-0.08$. This revision implies a considerably shorter circularization timescale of $\tau_e\sim20\,$Myr, which in turn translates to a lower estimate of the tidal quality factor ($Q\sim10^4$ for our adopted parameters). We further note that we have assumed coplanarity within the system, and while a null mutual inclination between planets is expected based upon the inferred early dynamical evolution of the GJ\,876\,b-c resonance \citep{LeePeale2002}, a significant orbital misalignment would further complicate the interpretation of GJ\,876\,d's non-zero eccentricity.

We conclude by pointing out that although the orbital architecture of the GJ\,876 system itself is a relatively rare outcome of planet formation, it is almost certainly non-unique. That is, both multi-planetary resonant chains (e.g. Kepler-60 and Kepler-223) as well as rapidly chaotic systems, such as Kepler-36 (\citealt{deck2012}) are present within the Galactic planetary census. This means that the dynamical calculation of tidal $Q$ presented herein is generally not limited to the GJ\,876 system, and will be applicable to other extrasolar systems as they continue to be discovered. In particular, an application of the derived methodology to transiting multi-planet systems would provide a viable avenue for further constraining the efficiency of tidal dissipation of extrasolar planets.

\section*{Acknowledgements}
We are grateful to Chris Spalding, Kat Deck, Elizabeth Bailey, Sarah Millholland and Greg Laughlin for illuminating discussions, as well as to the anonymous referee for providing an insightful report. This research was supported by NSF grant AST1517936 and Caltech's Summer Undergraduate Research Fellowship (SURF) program.

\bibliographystyle{apa}

\end{document}